\documentclass[a4paper,11pt]{article}
\usepackage{pos}

\newcommand{\apj}{Astrophysical J.}
\newcommand{\apjl}{Astrophysical J. Letters}

\newcommand{\apjs}{Astrophysical J. Supplement}
\newcommand{\aap}{Astronomy and Astroph.}

\newcommand{\aapr}{Astronomy and Astroph. Reviews}
\newcommand{\mnras}{MNRAS}
\newcommand{\pasa}{Publ. Astron. Soc. of Australia}
\newcommand{\araa}{Annual Review of Astronomy and Astroph.}

\newcommand{\nat}{Nature}

\manuallySeparateAuthors

\title{Hunting the gamma-ray emission from Fast Radio Burst with Fermi-LAT}
 \ShortTitle{Gamma-ray emission from FRBs with Fermi-LAT}

\author*[a,b,c]{Giacomo Principe}
\author[d]{, Niccolò Di Lalla}
\author[e]{, Leonardo Di Venere}
\author[f]{, Michela  Negro}
\author[a,b]{, Francesco Longo}
\author{, on behalf of the Fermi Large Area Telescope Collaboration.}

\affiliation[a]{Universit\'a di Trieste, Dipartimento di Fisica,\\  I-34127 Trieste, Italy}

\affiliation[b]{Istituto Nazionale di Fisica Nucleare, Sezione di Trieste,\\ I-34127 Trieste, Italy}

\affiliation[c]{Istituto Nazionale di Astrofisica - Istituto di Radioastronomia,\\ I-40129 Bologna, Italy}

\affiliation[d]{W. W. Hansen Experimental Physics Laboratory, Kavli Institute for Particle Astrophysics and Cosmology, Department of Physics and
SLAC National Accelerator Laboratory, Stanford University,\\  94305, Stanford, CA, USA}

\affiliation[e]{Istituto Nazionale di Fisica Nucleare, Sezione di Bari,\\ I-70126 Bari, Italy}

\affiliation[f]{University of Maryland, Baltimore County, Baltimore, MD 21250, USA - NASA Goddard Space Flight Center, Greenbelt, MD 20771, USA - Center for Research and Exploration in Space Science and Technology, NASA/GSFC, Greenbelt, MD 20771, USA}

\emailAdd{giacomo.principe@ts.infn.it}

\abstract{Fast radio bursts (FRBs) are one of the most exciting new mysteries of astrophysics. Their origin is still unknown, but recent observations seem to link them to soft gamma repeaters and, in particular, to magnetar giant flares (MGFs). The recent detection of a MGF at GeV energies by the Fermi Large Area Telescope (LAT) motivated the search for GeV counterparts to the >100 currently known FRBs. To date, none of these has a known gamma-ray counterpart.
Taking advantage of more than 12 years of Fermi-LAT data, we perform a search for gamma-ray emission from almost all the reported repeating and non-repeating FRBs. We analyze on different time scales the Fermi-LAT data for each individual source separately and perform a cumulative analysis on the repeating ones. In addition, we perform the first stacking analysis at GeV energies of this class of sources in order to constrain the gamma-ray properties of the FRBs. The stacking analysis is a powerful method that allows for a possible detection from below-threshold FRBs providing important information on these objects. In this proceeding we present preliminary results of our study and we discuss their implications for the predictions of gamma-ray emission from this class of sources.}

\FullConference{%
  7th Heidelberg International Symposium on High-Energy Gamma-Ray Astronomy (Gamma2022)\\
  4-8 July 2022\\
  Barcelona, Spain\\}

\begin{document}
\maketitle

\section{Introduction}
Fast Radio Burst (FRBs) are bright (typical fluences of few Jy), short-duration (few ms or less) radio pulses that flash randomly in the sky \citep{2019A&ARv..27....4P}. They are thought to be caused by some high-energy astrophysical processes not yet understood.
FRBs are expected to have extragalactic origin due their high dispersion measure (DM, the free electron density along the line of sight) in excess to the Galactic value \citep{2019ARA&A..57..417C}. 

More than 1000 FRBs have been reported as of today \citep[e.g.; see FRB catalogs][]{2016PASA...33...45P,2021ApJS..257...59C}. Among them, the number of FRBs presenting repeating events is increasing More than 1000 FRBs have been reported as of today \citep[e.g.; see FRB catalogs][]{2016PASA...33...45P,2021ApJS..257...59C}. Among them, the number of FRBs presenting repeating events is increasing over time in recent years, currently more than 20 repeaters have been detected.

Despite being discovered more than 15 years ago \citep{2007Sci...318..777L}, their origin is still unknown \citep{2021Univ....7...76N}. In April 2020, a FRB-like emission was first associated with the Galactic magnetar SGR\,1935+2154, the soft gamma repeater \citep{2020Natur.587...54C}, thanks to a coincident X-ray burst recorded by the INTEGRAL and AGILE telescopes \citep[][, respectively]{2020ApJ...898L..29M,2021NatAs...5..401T}. 
This event supported the scenario that magnetars can be the progenitors of FRBs.
Recently (October 2022), a few more coincident radio and high energy events
were detected from the same source \citep[see e.g.][]{2022ATel15686....1F}.
Assuming that all FRBs are produced by magnetar flares, we expect that all FRBs should be accompanied by an afterglow emission peaking at gamma-ray energies (E$_{\textrm{peak}} >$ MeV–GeV), with energies $\eta \geq 10^{4}$ larger than the emitted radio energy \citep{2020ApJ...899L..27M}.

The detection, in April 2020, of high-energy emission (up at GeV energies) from a magnetar giant flare in the nearby Sculptor galaxy by \citep{2021NatAs...5..385F} motivated our search for gamma-ray counterparts to the known FRBs.
Few searches for FRBs counterparts at gamma-ray energies have been performed in the latest years without any significant detection \citep[see e.g,][]{2019ApJ...879...40C,2020A&A...637A..69G,2020ApJ...893L..42T,2021ApJ...915..102V}, but only analysing a limited sample of FRBs (up to a few dozens).
Taking advantage to observations done by the \textit{Fermi} Large Area Telescope (LAT), with more than 13 years of data collected, and to more than 1000 published FRBs events, we aim to perform a large and deep systematic search for high-energy counterparts of the reported repeating and non-repeating bursts.  
Throughout this proceeding, we assume $H_{0} = 70$ km s$^{-1}$ Mpc$^{-1}$ , $\Omega_{M} = 0.3$, and $\Omega_{\Lambda}=0.7$ in a flat Universe.

\section{Sample of FRBs}
Our FRBs sample consists of 1025 events, including 561 non-repeating FRBs and 459 bursts from 22 repeaters. The sample has been obtained selecting the events from the following resources:
\begin{itemize}
    \item 118 FRBs collected in the FRBCAT\footnote{https://frbcat.org} \citep{2016PASA...33...45P};
    \item 535 repeating and non-repeating FRBs reported in the first CHIME/FRB catalog \citep{2021ApJS..257...59C};
    \item 230 events from the 20 repeaters in the 1st CHIME-FRB catalog, CHIME/FRB collaboration  (http://www.chime-frb.ca/repeaters) listing the new events for these sources, as of June 15, 2021,
    \item 235 bursts from FRB\,121102 collected by \citet{2020MNRAS.495.3551R}.
\end{itemize}

Among them, only fourteen have been detected with arcsec precision enabling the identification of their host galaxies \footnote{http://frbhosts.org/}, which have luminosity distances ranging from 3.6 Mpc to 4 Gpc, giving solid foundations to their cosmological origin.
In addition, the FRB-like emission seen on April 28, 2020 has been associated to a Galactic magnetar giant flares (MGFs), SGR 1935+2154, located at a distance of about 14 kpc \citep[FRB 200428,][]{2020Natur.587...59B}.

\begin{figure}
\centering
\includegraphics[width=14cm]{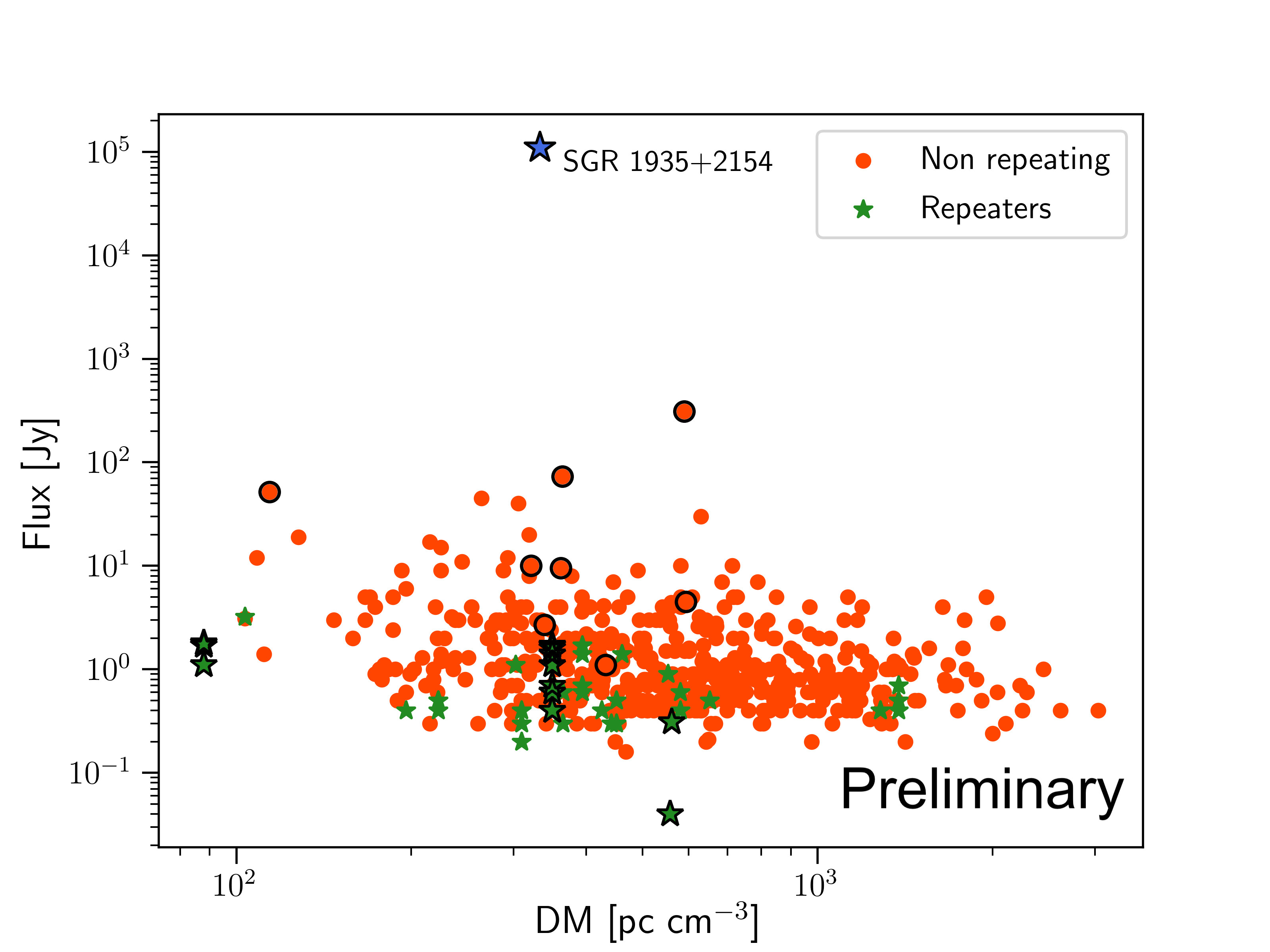}
\caption{\small \label{fig:flux_dm}
Diagram of the radio flux vs DM distribution for the FRBs contained in our sample. The FRBs with known host galaxies are highlighted with a black contour.}
\end{figure}

Fig. \ref{fig:flux_dm} shows the diagram of the radio flux as a function of DM for FRBs in our sample. They present a DM value ranging from 87 to 3000 pc cm$^{-3}$. Their large dispersion measure supports their possible extragalactic origin. 
In particular, 14 FRBs contained in our sample present an accurate radio localisation pointing to their host galaxy.
The FRBs present a radio flux between $\sim$ 0.05 Jy and over 100 kJy. The highest flux reported for the FRBs in our sample has been recorded for the event associated to the Galactic magnetar SGR\,1935+2154.

\noindent More information on the selected sample have been provided in \citet{2022icrc.confE.624P}, and they will be better described in a forthcoming publication.

\section{\textit{Fermi}-LAT analysis}
Fermi-LAT is a gamma-ray telescope which collects photons using the electron-positron pairs effect. It has an operational energy range from 20\,MeV to $\sim$\,TeV. The LAT is composed by a high-resolution converter tracker (for direction measurement of the incident gamma rays), a CsI(Tl) crystal calorimeter (for energy measurement) and an anti-coincidence detector to identify the background of charged particles \citep{2009ApJ...697.1071A}. Thanks to its wide-field of view (covering the entire sky in $\sim 3$ hours), the LAT represents a suitable $\gamma$-ray detector for searching for the high-energy counterpart of FRBs.

In order to deeply investigate the possible gamma-ray emission from FRBs we used various analysis techniques on different time scales ranging from a few seconds up to several years.
In addition, we performed a stacking analysis on the entire sample of events, in order to search for cumulative emission \citep[a similar method was applied in ][ to study another class of celestial objects]{2021MNRAS.507.4564P}.

In particular, for the results reported here we analysed each individual burst using time intervals of 10 and 10000 s.
The binned likelihood analysis (which consists of model optimisation, and localisation, spectrum and variability analyses) was performed with Fermipy\footnote{http://fermipy.readthedocs.io/en/latest/} \citep{2017ICRC...35..824W}, a python package that facilitates the analysis of LAT data with the \textit{Fermi} Science Tools (v. 11-07-00 was).
For our analysis we selected photons which have been reprocessed with the P8R3\_Source\_V2 instrument response functions (IRFs) \citep[IRFs,][]{2018arXiv181011394B}, in the energy range between 100\,MeV and 1\,TeV. The lowenergy threshold is motivated by the poor effective area and large uncertainties in the arrival directions of the photons below 100 MeV, leading to a possible confusion between point-like sources and the Galactic diffuse component \citep{2018A&A...618A..22P}.
For more details on the description of the analysis see \citet{2022icrc.confE.624P}.

\section{Preliminary \textit{Fermi}-LAT results on 10 and 10000 s}
We present here preliminary results focusing on the analysis of the 10 s and 10000 s time scales.
Among the FRBs in our sample, 258 events (958 events) have enough photons to correctly compute the likelihood analysis on the time interval of $\Delta_T$=10 s ($\Delta_T$=10000 s).
In addition, six FRBs were not considered in our analysis because they happened before the launch of the \textit{Fermi}-LAT satellite on June 2008 (namely FRB010724, FRB010621, FRB010312, FRB010305) or occurred during the end of March 2018, when the LAT was in safety mode for more than 23 days without taking any data due to a problem with a solar panel orientation (namely FRB180324, FRB180321) \citep{2021ApJS..256...12A}.

\begin{figure}[h]
\centering
\includegraphics[width=0.75\columnwidth]{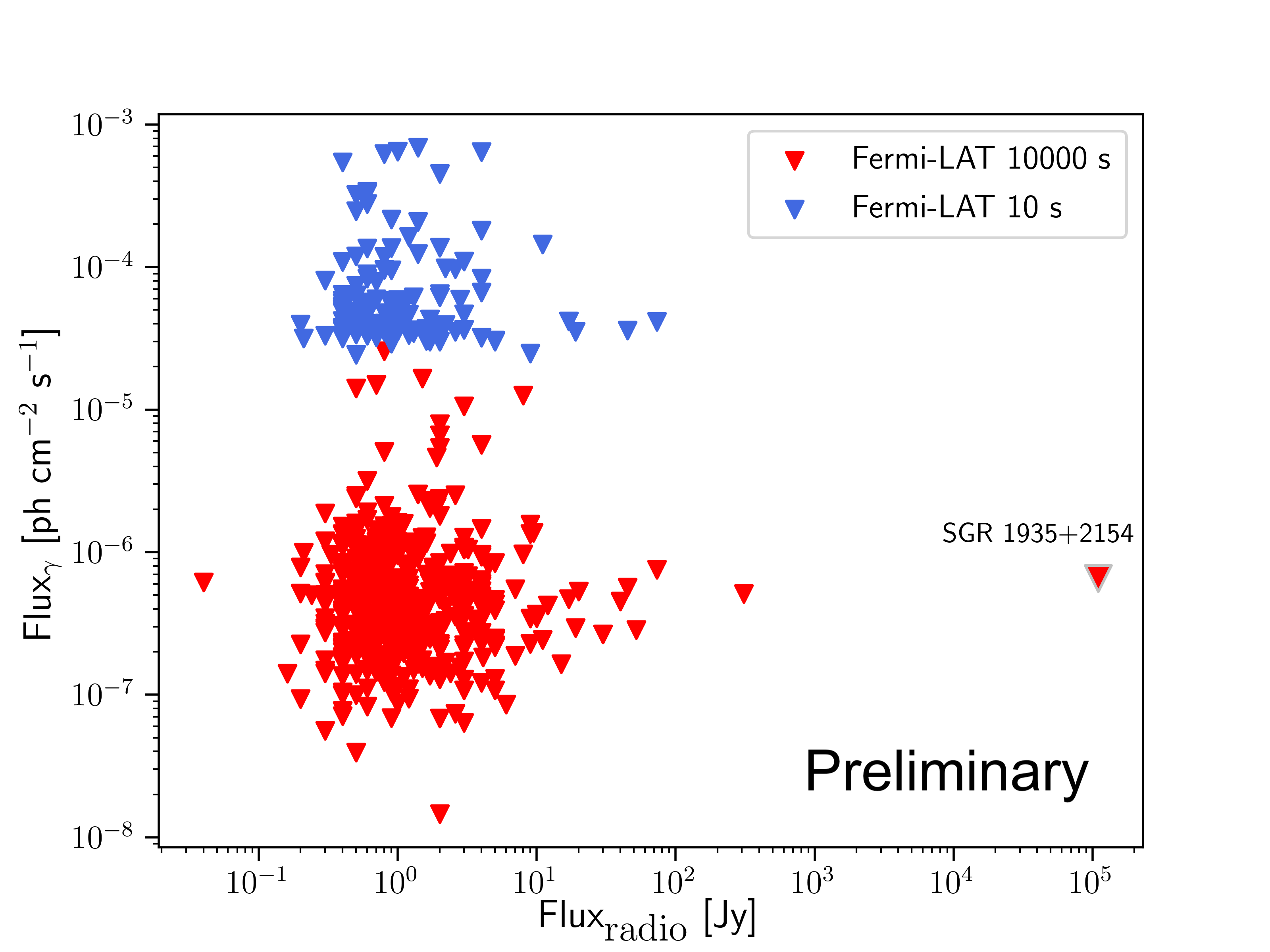}
\caption{\small \label{fig:flux_ul_fermi} 
Diagram with the upper limits on the gamma-ray vs. radio fluxes for the analysis on 10 s (in blue) and 10000 s (in red). The FRBs plotted here are those with sufficient gamma-ray exposure to perform the LAT analysis. Gamma-ray upper limits from the analysis on 10 s (10000 s) are plotted in blue (red).
}
\end{figure}

Figure \ref{fig:flux_ul_fermi} shows preliminary results with the upper limits (UL) on the gamma-ray flux compared to their measured radio flux.

The most stringent UL on the gamma-ray to radio flux ratio have been obtained for the event of SGR\,1935+2154 ($< 10^{-11}$ ph cm$^{-2}$ s$^{-1}$ Jy$^{-1}$).

\section{Conclusion}
FRBs are a recent class of sources which represents one of the most intriguing questions in astrophysics. Despite more than a thousand of detections, their origin is still unclear.
They are expected to emit gamma-ray up to energies of hundreds of GeV \citep{2019MNRAS.485.4091M}.
Thanks to a broad sample of FRBs (more than 1000) and different analysis techniques we aim to investigate their gamma-ray emission.
Preliminary results on 10 s and 10000 s time intervals around the FRBs episode do not show any significant emission, providing anyhow stringent UL on the gamma-ray to radio flux ratio ($< 10^{-11}$ ph cm$^{-2}$ s$^{-1}$ Jy$^{-1}$).
These results will provide important information for better constraining the FRB origin and modelling the emission mechanisms at high energies.
Further more complete results on our search for the high-energy emission from FRBs will be reported in a forthcoming publication \citep{principe_in_preparation}. 

\subsection*{ACKNOWLEDGMENTS}
We acknowledge use of the CHIME/FRB Public Database, provided at https://www.chime-frb.ca/ by the CHIME/FRB Collaboration.

The \textit{Fermi}-LAT Collaboration acknowledges support from NASA and DOE (United States), CEA/Irfu, IN2P3/CNRS, and CNES (France), ASI, INFN, and INAF (Italy), MEXT, KEK, and JAXA (Japan), and the K.A. Wallenberg Foundation, the Swedish Research Council, and the National Space Board (Sweden).

%
%
%

\end{document}